\documentclass[twocolumn, showpacs,prl, floatfix, ]{revtex4}
\usepackage{dcolumn}
\usepackage{bm}
\usepackage{amsmath}
\usepackage{amsfonts}
\usepackage{amssymb}

\input{tcilatex}
\begin{document}

\title{Control and Representation of n-qubit Quantum Systems }
\author{W. E. Baylis, R. Cabrera, C. Rangan}
\affiliation{Department of Physics, University of Windsor, Windsor, ON N9B 3P4. Canada}

\begin{abstract}
Just as any state of a single qubit or 2-level system can be obtained from
any other state by a rotation operator parametrized by three real Euler
angles, we show how any state of an $n$-qubit or $2^{n}$-level system can be
obtained from any other by a compact unitary transformation with $2^{n+1}-1$
real angles, $2^{n}$ of which are azimuthal-like and the rest polar-like.
The results follow from a modeling of the Hilbert space of $n$-qubits by a
minimal left ideal of an associative algebra. This representation is
expected to be useful in the design of new compact control techniques or
more efficient algorithms in quantum computing.
\end{abstract}

\pacs{03.67.-a, 03.65.Fd, 03.67.Mn, 02.20.Sv }
\maketitle

The general control of an $N$-level quantum system is a subject of
considerable interest in many fields, including chemical dynamics and
quantum computing. A general statement that establishes the controllability
basically says that for an $N$-level system to be completely controllable,
it is sufficient that the free-evolution Hamiltonian, along with the
interaction Hamiltonian (which could involve a sequence of steps) and all
possible commutators among them, form a Lie algebra, which in general is $%
u(N),$ with $N^{2}$ independent real elements~\cite{Brockett1972,Ramakrishna}%
. Once quantum controllability is established, it is important to find the
optimal method to actually control the system. This involves optimization of
both resources and time (for examples, see \cite%
{Geremia2004,KhanejaBrockett,RanganPRA2001} and references therein).

In this letter, we show that an arbitrary state of an $N$-level quantum
system can be generated from any other state by a compact, rotation-like
unitary transformation with $2N-1$ real parameters. We explicitly
demonstrate this controllability for $n$-qubit systems where $N=2^{n}$. We
show that a complex algebra of only $N$ independent elements is sufficient
to describe an evolution from any given reference state to an arbitrary
state of the system. The $N=2^{n}$ independent elements of the linear space
of the algebra span the $N$-dimensional Hilbert space, and the $N-1$
independent elements excluding unity generate polar-like rotations. When
combined with the $N$ independent phase angles for each term, which arise
from azimuthal-like rotations, there are a total of $2^{n+1}-1=2N-1$ real
control parameters for the system. Thus, it is shown that the minimum number 
$2N-1$ of real parameters needed to identify an arbitrary state is also
sufficient to create it from any other state by a compact, rotation-like
unitary operation. Our results generalize the well-known description of a
single-qubit system, any state of which can be specified by the three Euler
angles of a rotation that relate the state to a \textquotedblleft
spin-up\textquotedblright\ reference state.

Clifford's geometric algebra has been used to study Lie groups without the
use of matrices \cite{CDoran}. The Clifford algebra $C\!\ell _{N}$ of an $N$%
-dimensional Euclidean space contains a scalar, $N$ linearly independent
vectors or directions, $\binom{N}{2}=N\left( N-1\right) /2$ linearly
independent bivectors or planes, $\binom{N}{3}$ linearly independent
trivectors, and so on for a total of $2^{N}$ linearly independent elements.
Its unimodular even elements, called rotors, form the Spin($N$) group, which
is a two-fold cover of the rotation group SO($N$) and is generated by its
bivectors. The 2-state spinor of a single qubit can be represented as the
projection of a rotor in the Clifford algebra $C\!\ell _{3}$ of
3-dimensional Euclidean space onto a minimal left ideal. \cite%
{Bay03a,Bay96,Bay99,Bay92}. An equivalent representation has been extended
to the Lie algebra of two-qubit systems \cite{Havel}. This letter shows how
this structure can be simply implemented for an $n$-qubit system, and more
generally, for an $N$-level quantum system.

The qubit or 2-state quantum system is well known in many contexts,
including the Bloch sphere for spin-1/2 magnetic resonance \cite{Bloch} and
the Feynman-Vernon-Hellwarth model \cite{FeynmanVH}, both of which
ultimately have the same mathematical structure. The states of a single
qubit can be represented by elements of Spin(3), which is isomorphic to
SU(2), projected onto a minimal left ideal of $C\!\ell _{3}$ by the
projector $P_{3}=\frac{1}{2}\left( 1+\sigma _{3}\right) $ with $\sigma _{3}$
a unit vector. Projectors are real idempotent elements that obey $%
P_{3}=P_{3}^{2}=\sigma _{3}P_{3}~.$ In a matrix representation of $C\!\ell
_{3}$, the unit vectors $\sigma _{j}$ can be represented by the Pauli-spin
matrices, $\sigma _{0}\equiv 1$ is represented by the unit matrix, and the
volume element is $\sigma _{1}\sigma _{2}\sigma _{3}=i1$. The projector acts
from the right to reduce the matrix representation of the rotor $R$ to a
spinor $\psi =RP_{3}$ with a single non-vanishing column. Within the minimal
left ideal, the basis states of the two-level system can be taken as 
\begin{eqnarray}
P_{3} &=&\sigma _{3}P_{3}\;\text{(\textquotedblleft spin
up\textquotedblright )} \\
\sigma _{1}P_{3} &=&\sigma _{13}P_{3}\;\text{(\textquotedblleft spin
down\textquotedblright )}  \notag
\end{eqnarray}%
where $\sigma _{13}=\exp \left( \pi \sigma _{13}/2\right) $ is the bivector $%
\sigma _{13}=\sigma _{1}\sigma _{3}=-\sigma _{3}\sigma _{1}$ that generates
rotations with rotor $R=\exp \left( \theta \sigma _{13}/2\right) $ in the
plane of $\sigma _{1}$ and $\sigma _{3},$ and we noted that $\sigma
_{3}^{2}=1=-\sigma _{13}^{2}.$ A general state is a linear combination%
\begin{equation}
\psi =\left( c_{0}+c_{1}\sigma _{1}\right) P_{3}  \label{psi1q}
\end{equation}%
with complex scalar coefficients $c_{0},c_{1}.$ If the state is normalized: $%
\left\vert c_{0}\right\vert ^{2}+\left\vert c_{1}\right\vert ^{2}=1,$ we can
write $c_{0}=e^{i\alpha }\cos \theta /2$ and $c_{1}=e^{i\beta }\sin \theta
/2 $ with real phase angles $\alpha ,\beta ,$ and express its spinor by%
\begin{eqnarray}
\psi &=&\left( e^{i\alpha }\cos \frac{\theta }{2}-e^{i\beta }\sigma
_{13}\sin \frac{\theta }{2}\right) P_{3} \\
&=&e^{i\alpha }\left( \cos \frac{\theta }{2}-e^{-i\phi }\sigma _{13}\sin 
\frac{\theta }{2}\right) P_{3},\;\phi =\alpha -\beta  \notag \\
&=&e^{-i\phi \sigma _{3}/2}e^{-\sigma _{13}\theta /2}e^{-i\chi \sigma
_{3}/2}P_{3},\ \chi =\phi +2\alpha  \notag
\end{eqnarray}%
The left factor multiplying $P_{3}$ is here the rotor for a rotation with
Euler angles $\left( \phi ,\theta ,\chi \right) $. Its three real parameters
are the polar angle $\theta $ and two azimuthal angles $\chi $ and $\phi .$
In terms of physical operators, $\phi $ can be a phase generated by free
evolution, $\theta $ a phase generated by an applied field, and $\chi $ one
generated by both free evolution and geometry.

In this paper, we extend this formalism to systems of $N$ states, and
represent such states by unitary elements of a compact rotor-like group with 
$2N-1$ real parameters (angles), projected onto a minimal left ideal, and,
most significant for the control of the system, we show that transformations
between arbitrary states are given simply by such rotor-like operators.

The proof involves a simple algebraic representation of the states. In the
case of a single qubit, any state is given by a direction in
three-dimensional space (on the Bloch sphere) together with an overall
complex phase. However as seen in (\ref{psi1q}), it is more simply
represented by a complex associative algebra $\mathcal{A}_{1}$ of the left
factor with just one spatial dimension times a projector $P_{3}$ outside the
algebra. Every element of $\mathcal{A}_{1}$ for a single qubit is a complex
linear combination of $\left\{ 1,\mathbf{e}_{1}\right\} $ and $\mathcal{A}%
_{1}$ can be identified with the Clifford algebra $C\!\ell _{1}.$ We show
below that an arbitrary state of a system of $n$ qubits can similarly be
represented by the minimal projection of a left-factor algebra $\mathcal{A}%
_{n}$ of $N=2^{n}$ independent elements generated by products of $n$
fundamental ones. The algebra $\mathcal{A}_{n}$ can be identified as the
Clifford algebra $C\!\ell _{n}$ over the complex field, but other algebras
are also possible and it may be simpler to use an abelian algebra. In all
cases, the left factor can be written as the product of compact rotor-like
unitary transformations that can be factored into $N-1$ polar-like rotors
and $N$ azimuthal-like ones. The generators of the polar-like rotations give
the basis states of the system and the polar-like angles themselves give the
amplitudes of the basis states in the state expansion, whereas the
azimuthal-like angles determine the complex phases of the basis-state
expansion and are related to the geometric phases that can be found for $N$%
-level systems by an interative procedure recently discussed by Uskov and Rau%
\cite{UskovRau05}. The total number of real angular parameters is thus
exactly the minimum $2N-1$ parameters needed to define any state of the
system.

The Hilbert space of an $n$-qubit system is spanned by tensor products of $n$
single-qubit states. A complete basis of states is $\left\{ \left\vert \ell
\right\rangle =\mathfrak{b}\left( \ell \right) \psi _{0},~\ell =0,\ldots
,N-1\right\} ,$ where the reference state $\psi _{0}=P=P_{3}\otimes
P_{3}\otimes \cdots \otimes P_{3}$ is a primitive projector, and $\mathfrak{b%
}\left( \ell \right) $ are the hermitian operators 
\begin{equation}
\mathfrak{b}\left( \ell \right) =\sigma _{1}^{\lambda _{n}}\otimes \sigma
_{1}^{\lambda _{n-1}}\otimes \sigma _{1}^{\lambda _{n-2}}\otimes \cdots
\otimes \sigma _{1}^{\lambda _{1}},
\end{equation}%
where each $\lambda _{j},~j=1,2,\ldots ,n$ is $0$ or $1,$ with $0$
corresponding here to \textquotedblleft spin up\textquotedblright\ and $1$
to \textquotedblleft spin down\textquotedblright . To be explicit, we take $%
\lambda _{j}$ to be the bits in the binary expression of $\ell .$ In
particular, $\mathfrak{b}\left( 0\right) =1$ because the binary number $0$
has all bits $\lambda _{k}=0.$ Thus, we denote a general state $\psi $ by%
\begin{equation}
\psi =\sum_{\ell =0}^{N-1}\left\vert \ell \right\rangle c_{l}=\sum_{\ell
=0}^{N-1}c_{l}\mathfrak{b}\left( \ell \right) \psi _{0}.  \label{psi}
\end{equation}%
The $\mathfrak{b}\left( \ell \right) ,~\ell \in \left[ 0,n\right] \cap 
\mathbb{Z},$ form a closed abelian group with $\left[ \mathfrak{b}\left(
\ell \right) \right] ^{2}=1.$ The inner product of states $\psi $ and $\psi
^{\prime }$ is given algebraically by $\left\langle \psi ^{\prime }|\psi
\right\rangle =N\left\langle \psi \psi ^{\prime \dag }\right\rangle _{0},$%
where $\left\langle x\right\rangle _{0}$ is the scalar part of $x,$ that is,
the part that is scalar in each factor of its tensor product. It is easily
verified that the basis states $\left\vert \ell \right\rangle $ are
orthonormal. Since $\psi $ is assumed to be normalized, $\left\langle \psi
|\psi \right\rangle =\sum_{\ell =0}^{N-1}\left\vert c_{\ell }\right\vert
^{2}=1.$ A direct power-series expansion verifies that%
\begin{equation}
\exp \left( -i\mathfrak{b}\theta /2\right) =\cos \frac{\theta }{2}-i%
\mathfrak{b}\sin \frac{\theta }{2},  \label{Euler}
\end{equation}%
and a product of $N-1$ such compact unitary factors for some set of $N-1$
real \textquotedblleft polar angles\textquotedblright\ $\theta _{\ell }$
gives%
\begin{equation}
\prod_{\ell =1}^{N-1}\exp \left[ -\frac{i}{2}\mathfrak{b}\left( \ell \right)
\theta _{\ell }\right] =\exp \left[ -\frac{i}{2}\sum_{\ell =1}^{N-1}%
\mathfrak{b}\left( \ell \right) \theta _{\ell }\right] =\sum_{\ell
=0}^{N-1}c_{\ell }\mathfrak{b}\left( \ell \right)  \label{prodexp}
\end{equation}%
with complex coefficients normalized so that%
\begin{equation}
\sum_{\ell =0}^{N-1}\left\vert c_{\ell }\right\vert ^{2}=1.  \label{norm}
\end{equation}%
The expression (\ref{prodexp}) is similar to the state expansion (\ref{psi}%
), but the phases are not necessarily the same. To adjust the phases, we
include $N$ \textquotedblleft azimuthal angles\textquotedblright\ in
rotations generated by%
\begin{equation}
\mathfrak{z}\left( \ell \right) =\sigma _{3}^{\lambda _{n}}\otimes \sigma
_{3}^{\lambda _{n-1}}\otimes \sigma _{3}^{\lambda _{n-2}}\otimes \cdots
\otimes \sigma _{3}^{\lambda _{1}},\ \ell \in \left[ 0,n\right] \cap \mathbb{%
Z}.
\end{equation}%
The operators $\mathfrak{z}\left( \ell \right) $ are analogous to the $%
\mathfrak{b}\left( \ell \right) $ but with $\sigma _{3}$ replacing $\sigma
_{1}$. Each $\mathfrak{z}\left( \ell \right) $ either commutes or
anticommutes with any given $\mathfrak{b}\left( \ell ^{\prime }\right) $ and
is projected to unity by $\psi _{0}=P.$ Thus, a product of rotors%
\begin{eqnarray}
R &=&\prod_{\ell =0}^{N-1}\exp \left[ -\frac{i}{2}\mathfrak{z}\left( \ell
\right) \phi _{\ell }\right] \prod_{\ell ^{\prime }=0}^{N-1}\exp \left[ -%
\frac{i}{2}\mathfrak{b}\left( \ell ^{\prime }\right) \theta _{\ell ^{\prime
}}\right] \\
&=&\exp \left[ -\frac{i}{2}\sum_{\ell =0}^{N-1}\mathfrak{z}\left( \ell
\right) \phi _{\ell }\right] \exp \left[ -\frac{i}{2}\sum_{\ell ^{\prime
}=1}^{N-1}\mathfrak{b}\left( \ell ^{\prime }\right) \theta _{\ell ^{\prime }}%
\right]  \notag
\end{eqnarray}%
when projected onto the minimal left ideal by $\psi _{0}$ gives a unique
state $\psi =R\psi _{0}$ (\ref{psi}) with arbitrary phase factors. Again in
terms of physical operators, the $\theta $'s are phases that arise due to
applied fields and $\phi $'s arise mainly due to free evolution, but
sometimes due to geometry.

The $N$ basis-state operators $\mathfrak{b}\left( \ell \right) $ are not
unique. We can alternatively choose elements generating the Clifford algebra 
$C\!\ell _{n}$ with%
\begin{eqnarray}
\mathbf{e}_{1} &=&1\otimes \cdots \otimes 1\otimes 1\otimes \sigma _{1}=%
\mathfrak{b}\left( 1\right) \mathfrak{z}\left( 0\right)  \notag \\
\mathbf{e}_{2} &=&1\otimes \cdots \otimes 1\otimes \sigma _{1}\otimes \sigma
_{3}=\mathfrak{b}\left( 2\right) \mathfrak{z}\left( 1\right) \\
\mathbf{e}_{m} &=&\mathfrak{b}\left( 2^{m-1}\right) \mathfrak{z}\left(
2^{m-1}-1\right) .  \notag
\end{eqnarray}%
These anticommute and square to $+1:$ $\mathbf{e}_{j}\mathbf{e}_{k}+\mathbf{e%
}_{k}\mathbf{e}_{j}=2\delta _{jk}~.$ Many other bases are possible in the
form $\mathfrak{B}\left( \ell \right) P=\mathfrak{b}\left( \ell \right) 
\mathfrak{z}\left( L_{\ell }\right) $ for the expansion (\ref{psi}). Since $%
\mathfrak{z}\left( L_{\ell }\right) \psi _{0}=\psi _{0},$ the basis states
are actually the same, but the difference in operators is important since
given any set $\left\{ \mathfrak{B}\left( \ell \right) \right\} $, one can
find linear combinations $\sum_{\ell =0}^{N-1}c_{l}\mathfrak{B}\left( \ell
\right) $ that are singular, that is noninvertible, and such nonunitary
cannot be expressed as a rotor. Such cases do not cause problems since other
operator sets can be found in which the linear combinations are not singular.

The strength of our approach is evident from the example of the two-qubit or
four-level system rewritten in standard matrix form (although the algebra
does not require matrices). Four-level systems have been previously studied
in the context of coupled qubits \cite{Rau2000,Rau2005,Khaneja,Vala}, spin
3/2 fermions \cite{Congjun}, and controllability \cite{Rice,RanganJMP2005}.
The general state is (\ref{psi}) with $N=4$ and 
\begin{eqnarray*}
\psi _{0} &=&P=P_{3}\otimes P_{3}=\frac{1}{4}\sum_{\ell =0}^{3}\mathfrak{z}%
\left( \ell \right) =\left( 
\begin{array}{cccc}
1 & 0 & 0 & 0 \\ 
0 & 0 & 0 & 0 \\ 
0 & 0 & 0 & 0 \\ 
0 & 0 & 0 & 0%
\end{array}%
\right) \\
\mathfrak{b}\left( 0\right) &=&\mathfrak{z}\left( 0\right) =1\otimes
1=\left( 
\begin{array}{cccc}
1 & 0 & 0 & 0 \\ 
0 & 1 & 0 & 0 \\ 
0 & 0 & 1 & 0 \\ 
0 & 0 & 0 & 1%
\end{array}%
\right) \\
\mathfrak{b}\left( 1\right) &=&1\otimes \sigma _{1}=\left( 
\begin{array}{cccc}
0 & 1 & 0 & 0 \\ 
1 & 0 & 0 & 0 \\ 
0 & 0 & 0 & 1 \\ 
0 & 0 & 1 & 0%
\end{array}%
\right) \\
\mathfrak{b}\left( 2\right) &=&\left( 
\begin{array}{cccc}
0 & 0 & 1 & 0 \\ 
0 & 0 & 0 & 1 \\ 
1 & 0 & 0 & 0 \\ 
0 & 1 & 0 & 0%
\end{array}%
\right) ,\ \mathfrak{b}\left( 3\right) =\left( 
\begin{array}{cccc}
0 & 0 & 0 & 1 \\ 
0 & 0 & 1 & 0 \\ 
0 & 1 & 0 & 0 \\ 
1 & 0 & 0 & 0%
\end{array}%
\right) .
\end{eqnarray*}%
The matrices for $\mathfrak{z}\left( \ell \right) $ are diagonal and are
similarly obtained. Each $\mathfrak{b}\left( \ell \right) $ and $\mathfrak{z}%
\left( \ell \right) $ is its own inverse, but linear combinations are not
necessarily invertible as we demonstrate below. However, the state $\psi $
is arbitrary:%
\begin{equation}
\sum_{\ell =0}^{3}c_{\ell }\mathfrak{b}\left( \ell \right) \psi _{0}=\left( 
\begin{array}{cccc}
c_{0} & c_{1} & c_{2} & c_{3} \\ 
c_{1} & c_{0} & c_{3} & c_{2} \\ 
c_{2} & c_{3} & c_{0} & c_{1} \\ 
c_{3} & c_{2} & c_{1} & c_{0}%
\end{array}%
\right) \psi _{0}=\left( 
\begin{array}{cccc}
c_{0} & 0 & 0 & 0 \\ 
c_{1} & 0 & 0 & 0 \\ 
c_{2} & 0 & 0 & 0 \\ 
c_{3} & 0 & 0 & 0%
\end{array}%
\right) .  \label{psiinb}
\end{equation}%
In terms of compact unitary factors%
\begin{equation*}
\psi =\exp \left[ -\frac{i}{2}\sum_{\ell =0}^{3}\phi _{\ell }\mathfrak{z}%
\left( \ell \right) \right] \exp \left[ -\frac{i}{2}\sum_{\ell ^{\prime
}=1}^{3}\theta _{\ell ^{\prime }}\mathfrak{b}\left( \ell ^{\prime }\right) %
\right] \psi _{0}.
\end{equation*}%
We can relate the angles $\theta _{\ell ^{\prime }},\phi _{\ell }$ to the
complex coefficients $c_{\ell }$ by explicit expansions (\ref{Euler}) while
noting relations such as $\mathfrak{b}\left( 1\right) \mathfrak{b}\left(
2\right) =\mathfrak{b}\left( 3\right) $:%
\begin{eqnarray*}
&&\exp \left[ -\frac{i}{2}\sum_{b^{\prime }=1}^{3}\theta _{b^{\prime }}%
\mathfrak{b}\left( b^{\prime }\right) \right] \\
&=&\left[ \cos \frac{\theta _{1}}{2}\cos \frac{\theta _{2}}{2}\cos \frac{%
\theta _{3}}{2}+i\sin \frac{\theta _{1}}{2}\sin \frac{\theta _{2}}{2}\sin 
\frac{\theta _{3}}{2}\right] \mathfrak{b}\left( 0\right) \\
&&-i\left[ \sin \frac{\theta _{1}}{2}\cos \frac{\theta _{2}}{2}\cos \frac{%
\theta _{3}}{2}-i\cos \frac{\theta _{1}}{2}\sin \frac{\theta _{2}}{2}\sin 
\frac{\theta _{3}}{2}\right] \mathfrak{b}\left( 1\right) +\cdots
\end{eqnarray*}%
with the $\mathfrak{b}\left( 2\right) $ and $\mathfrak{b}\left( 3\right) $
coefficients obtained from that for $\mathfrak{b}\left( 1\right) $ by cyclic
permutation. This, plus the normalization condition (\ref{norm}) is
sufficient to relate the three polar angles $\theta _{j},~j=1,2,3,$ to the
magnitudes of the complex coefficients $c_{\ell }.$ The pattern of commuting
and anticommuting pairs $\mathfrak{z}\left( \ell \right) \mathfrak{b}\left(
\ell ^{\prime }\right) $ plus the relation $\mathfrak{z}\left( \ell \right)
P=P,$ implies%
\begin{equation*}
\exp \left[ -\frac{i}{2}\sum_{\ell ^{\prime }=0}^{3}\phi _{\ell ^{\prime }}%
\mathfrak{z}\left( \ell ^{\prime }\right) \right] b\left( \ell \right)
P=b\left( \ell \right) \exp \left( -\frac{i}{2}\alpha _{\ell }\right) P
\end{equation*}%
with%
\begin{eqnarray*}
\alpha _{0} &=&\phi _{0}+\phi _{1}+\phi _{2}+\phi _{3} \\
\alpha _{1} &=&\phi _{0}-\phi _{1}+\phi _{2}-\phi _{3} \\
\alpha _{2} &=&\phi _{0}+\phi _{1}-\phi _{2}-\phi _{3} \\
\alpha _{3} &=&\phi _{0}-\phi _{1}-\phi _{2}+\phi _{3}
\end{eqnarray*}%
and this is enough to relate the complex phases of the $c_{\ell }$ to the
four azimuthal angles $\phi _{k},~k=0,1,2,3.$

As mentioned above, for any choice of basis operators $\mathfrak{B}\left(
\ell \right) ,$ there exist some states for which the linear combination $%
\sum_{\ell }c_{\ell }\mathfrak{B}\left( \ell \right) $ is singular, that is,
has no inverse. For the basis operators $\mathfrak{b}\left( \ell \right) ,$
for example, the problem arises with the entangled states $\psi _{\pm }=%
\frac{1}{\sqrt{2}}\left( \mathfrak{b}\left( 0\right) \pm \mathfrak{b}\left(
3\right) \right) \psi _{0}~.$It is seen by inspection that the matrix $%
\sum_{\ell =0}^{3}c_{\ell }\mathfrak{b}\left( \ell \right) $ given in (\ref%
{psiinb}) has a vanishing determinant when $c_{3}=\pm c_{0}$ and $%
c_{1}=0=c_{2}.$ The problem is resolved by using different basis operators.
Thus, if we adopt the $C\!\ell _{2}$ basis with phases that make all basis
elements antihermitian, we get the expansion%
\begin{equation}
\sum_{\ell =0}^{3}c_{\ell }\mathfrak{B}\left( \ell \right) =\left( 
\begin{array}{rrrr}
c_{0} & ic_{1} & ic_{2} & -c_{3} \\ 
ic_{1} & c_{0} & c_{3} & -ic_{2} \\ 
ic_{2} & -c_{3} & c_{0} & ic_{1} \\ 
c_{3} & -ic_{2} & ic_{1} & c_{0}%
\end{array}%
\right) ,
\end{equation}%
which is unitary for the choice $c_{0}=c_{3}=1/\sqrt{2}$ and $c_{1}=0=c_{2}$.

The method can be applied to arbitrary $2^{n}$-level systems. The relation
of the tensor-product basis states, interpreted as giving spin orientations
of separate spin-1/2 systems, to the eigenstates of the system depends on
the Hamiltonian. For example, for $n=2,$ the Hamiltonian%
\begin{equation}
H=E_{0}+\gamma _{1}\mathbf{s}_{1}\cdot \mathbf{B}\otimes 1+1\otimes \gamma
_{2}\mathbf{s}_{2}\cdot \mathbf{B}+2\lambda \mathbf{s}_{1}\cdot \mathbf{s}%
_{2},
\end{equation}%
where $\mathbf{B}$ is the external magnetic field, $\gamma _{j}$ are the
gyromagnetic ratios, and $2\lambda $ is the spin-spin coupling constant, has
the matrix form%
\begin{equation}
\left( 
\begin{array}{cccc}
\hbar \omega _{0++}+\frac{\lambda }{2} & 0 & 0 & 0 \\ 
0 & \hbar \omega _{0+-}-\frac{\lambda }{2} & \lambda & 0 \\ 
0 & \lambda & \hbar \omega _{0-+}-\frac{\lambda }{2} & 0 \\ 
0 & 0 & 0 & \hbar \omega _{0--}+\frac{\lambda }{2}%
\end{array}%
\right) ,
\end{equation}%
where $\omega _{0\pm \pm }\equiv \omega _{0}\pm \omega _{1}\pm \omega _{2}.$
States $\psi _{0}=P$ and $\psi _{3}=\mathfrak{b}\left( 3\right) \psi
_{0}=\sigma _{3}\otimes \sigma _{3}P$ are eigenstates, but linear
combinations of $\psi _{1}=\mathfrak{b}\left( 1\right) P$ and $\psi _{2}=%
\mathfrak{b}\left( 2\right) P$ form the other eigenstates. In the 2-D basis $%
\left\{ \psi _{1},\psi _{2}\right\} ,$%
\begin{equation}
H=\hbar \omega _{0}-\lambda /2+\left( 
\begin{array}{cc}
\Delta & \lambda \\ 
\lambda & -\Delta%
\end{array}%
\right)
\end{equation}%
with $\Delta =\hbar \left( \omega _{1}-\omega _{2}\right) $ and the
eigenenergies are $\hbar \omega _{0}-\lambda /2\pm \sqrt{\Delta ^{2}+\lambda
^{2}}.$Since we have a total of four independent energy parameters, we
expect to be able to represent an arbitrary 4-level system with these
coupled qubits.

More generally, however, there is no need to associate the tensor-product
states literally with independent spins on spin-$\frac{1}{2}$ particles.
Tensor products of Pauli spin matrices are simply a way to specify matrix
representations of orthogonal states and operators on them. A complete set
of $2^{n}\times 2^{n}$ matrices is always given by tensor products of $n$
spin matrices $\sigma _{\mu }.$ Nothing inherent in our treatment restricts
it to systems of $n$-qubits. Any set of $N$ levels can be similarly handled.
In the same vein, the reference state $\psi _{0}$ need not be one with
maximum total spin component along the $z$ axis. Formally, we can transform
the reference state multiplying our spinors $\psi $ by a transformation
rotor from the right. This shows that the transformation between any two
states of the system can be represented by a unitary rotor with $N-1$
polar-like angles and $N$ azimuthal ones. For $N>1,$ this is fewer than the $%
N^{2}$ real parameters required to specify a unitary transformation on the
system in general, but this reduction simply reflects the fact that only a
small subset of the possible unitary transformations are required to
transform between arbitrary states. For most practical problems in quantum
control, it is only necessary to transform a few given states to others (for
example, making a quantum gate), and therefore our representation can be
expected to lead to more compact control schemes.

In summary, we have demonstrated that evolution of an arbitrary $n$-qubit or 
$N=2^{n}$-level quantum state into another state can be described by a
complex algebra of only $N$ independent elements. This compact unitary
transformation consists of $2^{n+1}-1$ real angles, $2^{n}$ of which are
azimuthal-like and $2^{n}-1$ that are polar-like. These results are expected
to lead to optimal schemes for coherent control and quantum computing.

It is pleasure to acknowledge helpful discussions with Prof. A.R.P. Rau and
Dr. D. B. Uskov of Louisiana State University. WEB and CR gratefully
acknowledge support of the research by the Natural Science and Engineering
Research Council of Canada.

\end{document}